\documentclass[aps,prd,showpacs,floatfix,nofootinbib,twocolumn,10pt]{revtex4}

\usepackage{amsmath,amssymb,graphicx,latexsym}
\usepackage{threeparttable,txfonts,subfigure,color}
\usepackage[colorlinks,linkcolor=blue,citecolor=blue,urlcolor=blue]{hyperref}

\def\aap{Astron.\ Astrophys.\ }

\def\apjl{Astrophys.\ J.\ Lett.\ }

\def\mnras{Mon.\ Not.\ Roy.\ Astron.\ Soc.\ }

\def\physrep{Phys.\ Rept.\ }
\def\prd{Phys.\ Rev.\ D\ }
\def\prl{Phys.\ Rev.\ Lett.\ }

\def\jcap{J.\ Cosmol.\ Astropart.\ Phys.\ }

\begin{document}

\title{Interpretations of the cosmic ray secondary-to-primary ratios 
measured by DAMPE}

\author{Peng-Xiong Ma$^{a}$}
\author{Zhi-Hui Xu$^{a,b}$}
\author{Qiang Yuan$^{a,b}$\footnote{yuanq@pmo.ac.cn}}
\author{Xiao-Jun Bi$^{c,d}$}
\author{Yi-Zhong Fan$^{a,b}$}
\author{Igor V. Moskalenko$^e$}
\author{Chuan Yue$^a$}

\affiliation{
$^a$Key Laboratory of Dark Matter and Space Astronomy, Purple Mountain
Observatory, Chinese Academy of Sciences, Nanjing 210023, China \\
$^b$School of Astronomy and Space Science, University of Science and
Technology of China, Hefei 230026, China\\
$^c$Key Laboratory of Particle Astrophysics, Institute of High Energy
Physics, Chinese Academy of Sciences, Beijing 100049, China\\
$^d$University of Chinese Academy of Sciences, Beijing 100049, China\\
$^e$W. W. Hansen Experimental Physics Laboratory and Kavli Institute for 
Particle Astrophysics and Cosmology, Stanford University, Stanford, 
CA 94305, USA
}

\begin{abstract}
Precise measurements of the boron-to-carbon and boron-to-oxygen ratios
by DAMPE show clear hardenings around $100$ GeV/n, which provide
important implications on the production, propagation, and interaction
of Galactic cosmic rays. In this work we investigate a number of models
proposed in literature in light of the DAMPE findings. These models
can roughly be classified into two classes, driven by propagation 
effects or by source ones. Among these models discussed, we find that the
re-acceleration of cosmic rays, during their propagation, by random 
magnetohydrodynamic waves may not reproduce sufficient hardenings of 
B/C and B/O, and an additional spectral break of the diffusion coefficient 
is required. The other models can properly explain the hardenings of the 
ratios. However, depending on simplifications assumed, the models differ 
in their quality in reproducing the data in a wide energy range. 
The models with significant re-acceleration effect will under-predict 
low-energy antiprotons but over-predict low-energy positrons, and the 
models with secondary production at sources over-predict high-energy
antiprotons. For all models high-energy positron excess exists.
\end{abstract}

\date{\today}


\maketitle

\section{Introduction}

Galactic cosmic rays (GCRs) are energetic particles produced by powerful
astrophysical objects such as the remnants of supernova explosions.
After being accelerated up to very high energies, they propagate and
interact in the Milky Way before entering the solar system and being
recorded by our detectors. There are typically two types of GCRs,
the primary family (such as protons, helium, carbon, oxygen, neon,
magnesium, silicon, and iron) which is produced directly by acceleration
at their sources and the secondary family (such as lithium, berylium, 
boron, and sub-iron nuclei) which is produced via fragmentations of primary 
particles mainly during the propagation process. Precise measurements
of the ratios between secondary particles and their parent primary
particles are important probe of the propagation of GCRs as well as
the turbulent properties of the interstellar medium (ISM)
\cite{1964ocr..book.....G,1990cup..book.....G,1990acr..book.....B}.

Among various secondary-to-primary ratios of nuclei, the boron-to-carbon
ratio (B/C) is the best measured and most widely studied. Measurements of 
B/C up to kinetic energies\footnote{In this paper, we are necessarily using 
the mixed energy units: discussions of the injection spectra and cosmic ray 
transport is done in terms of rigidity, while a comparison with experiments 
requires a conversion to the kinetic energy per nucleon.} of hundreds of 
GeV/n have been achieved with good precision by many experiments
\cite{1990A&A...233...96E,1990ApJ...349..625S,1991A&A...247..163F,2008ICRC....2....3P,2008APh....30..133A,2009ApJ...698.1666G,2010ApJ...724..329A,2011ApJ...742...14O,2014ApJ...791...93A,2016ApJ...831...18C,2016PhRvL.117w1102A,2019AdSpR..64.2559G,2018PhRvL.120b1101A},
which were extensively used to constrain the propagation of GCR models
(e.g., \cite{1991ApJ...374..356M,1998ApJ...509..212S,2001ApJ...555..585M,
2009ApJ...697..106A,2010A&A...516A..66P,2011ApJ...729..106T,
2012ApJ...752...69O,2015JCAP...09..049J,2016ApJ...824...16J,
2016PhRvD..94l3019K,2016PhRvD..94l3007F,2017PhRvD..95h3007Y,
2018PhRvD..97b3015N,2019PhLB..789..292W}).
The B/C ratio above $O(10)$ GV can be well fitted by a power-law function
of rigidity, $\propto{\mathcal R}^{-\delta}$, with $\delta\approx 1/3$
\cite{2016PhRvL.117w1102A}, in agreement with the prediction of GCR diffusion 
in the ISM with a Kolmogorov type turbulence spectrum 
\cite{1941DoSSR..30..301K,1998ApJ...509..212S}.
Further measurements of ratios of secondary lithium, beryllium, and 
boron to primary carbon and oxygen by AMS-02 jointly showed a hardening
\cite{2018PhRvL.120b1101A,2021PhR...894....1A}. 
Non-trivial spectral shapes of the secondary-to-primary ratios thus
challenge the simple production and propagation models of GCRs.

Very recently, high-precision measurements up to 5 TeV/n of the boron-to-carbon
(B/C) and boron-to-oxygen (B/O) ratios have been obtained by the Dark Matter 
Particle Explorer (DAMPE; \cite{ChangJin:550,2017APh....95....6C}). The DAMPE 
results revealed clear hardening of both ratios with high significance at 
nearly the same kinetic energy of $\sim 100$ GeV/n \cite{2022DAMPE-BC}. 
A broken power-law fit to the B/C (B/O) ratio gives a low-energy slope of 
0.356 (0.394) and a high-energy slope of 0.201 (0.187), and the change of 
slope is $\Delta\gamma=0.155$ (0.207). Previous measurements showed also 
remarkable hardenings of primary nuclei at similar energies 
\cite{2009BRASP..73..564P,2010ApJ...714L..89A,2017PhRvL.119y1101A,
2020PhRvL.125y1102A,2011Sci...332...69A,2019SciA....5.3793A,
2021PhRvL.126t1102A}. The slope changes of primary nuclei are about 
$0.1\sim0.2$, which are slightly diverse among different measurements.
These spectral features of GCRs may suggest a common origin.

A straightforward interpretation of the hardenings of B/C and B/O is the 
existence of a break of the diffusion coefficient at a few hundred GV 
\cite{2012ApJ...752...68V,2020ApJS..250...27B,2017PhRvL.119x1101G}. 
Such a break of the diffusion coefficient may be a consequence of the 
change of the scale-dependence of the ISM turbulence, or be due to the 
nonlinear particle-wave interactions \cite{2012PhRvL.109f1101B}. Other 
interpretations with different physical models were also proposed (e.g.,
\cite{2016ApJ...827..119C,2018PhRvD..97f3008G,2019MNRAS.488.2068B,
2020JCAP...11..027Y,2022ApJ...933...78M,2021PhRvD.104j3029M,
2021ApJ...917...61K}). These models either employ more complicated propagation 
effects or introduce additional sources of (secondary and/or primary) GCRs 
beyond the standard paradigm. 
Some of the above possibilities have been briefly discussed in 
Ref.~\cite{2022DAMPE-BC}. In this work we further explore these models to 
test whether they can explain the DAMPE data satisfactorily. Antiprotons 
and positrons from these models will also be discussed as independent 
tests of the models.

\section{Production and propagation model of Galactic cosmic rays}

The propagation of GCRs in the Milky Way can be generally described
by the diffusion equation
\begin{eqnarray}
\frac{\partial \psi}{\partial t} & = & 
\nabla\cdot(D_{xx}\nabla \psi-{\boldsymbol V_c}\psi)
+\frac{\partial}{\partial p}p^2D_{pp}\frac{\partial}
{\partial p}\frac{1}{p^2}\psi \nonumber \\
 & - & \frac{\partial}{\partial p}\left[\dot{p}\psi
-\frac{p}{3}(\nabla\cdot{\boldsymbol V_c}\psi)\right]
-\frac{\psi}{\tau_f}-\frac{\psi}{\tau_r}+q({\boldsymbol r},p)
\ , \label{prop}
\end{eqnarray}
which includes also the possible convective transportation effect with
velocity $\boldsymbol V_c$, the re-accerlation effect described by a
diffusion in the momemtum space with diffusion coefficient $D_{pp}$,
the energy losses with rate $\dot{p}$ and adiabatic losses,
fragmentations with time scale $\tau_f$, and radioactive decays with
lifetime $\tau_r$ \cite{2007ARNPS..57..285S}. The source function
$q({\boldsymbol r},p)$ includes both the primary contribution from
acceleration sources and the secondary contribution from GCR
interactions with the ISM.

The geometry of the propagation halo is assumed to be cynlindrially
symmetric, with radial extension $R_h=20$ kpc and height $\pm z_h$ to
be determined by the data. The spatial diffusion coefficient is usually
assumed to be spatially homogeneous, and depends on particle rigidity
with a power-law form
\begin{equation}
D_{xx}(\mathcal R)=D_0\beta^{\eta}\left(\frac{{\mathcal R}}
{{\mathcal R}_0}\right)^{\delta},\label{Dxx}
\end{equation}
where $\beta$ is the velocity of the particle in unit of light speed,
${\mathcal R}_0\equiv4$ GV is a reference rigidity, $\delta$ is the
power-law index describing the properties of the interstellar turbulence.
A phenomenological parameter $\eta$ is introduced to modify the velocity
dependence at low energies, in order to better match the measurements. 
We will discuss alternative cases about the
diffusion coefficient in this work (see below Sec. III for details).
The convection effect is neglected in this work according to the fitting
to the up-to-date data on GCR primary and secondary nuclei
\cite{2017PhRvD..95h3007Y,2019SCPMA..6249511Y}. The momentum diffusion
coefficient can be expressed as \cite{1994ApJ...431..705S}
\begin{equation}
D_{pp}=\frac{4p^2v_A^2}{3\delta(4-\delta^2)(4-\delta)wD_{xx}},
\label{Dpp}
\end{equation}
where $v_A$ is the Alfven speed of magnetized disturbances, $w$ is the ratio 
of magnetohydrodynamic (MHD) wave energy density to the magnetic field energy 
density and can be effectively absorbed into $v_A$.

The injection spectrum is assumed to be a smoothly broken power-law function
of rigidity
\begin{equation}
q({\mathcal R})=q_0{\mathcal R}^{-\gamma_0}\prod_{i=1}^n\left[1+\left(\frac
{\mathcal R}{{\mathcal R}_{{\rm br},i}}\right)^s\right]^{(\gamma_{i-1}-\gamma_i)/s},
\label{eq:inj}
\end{equation}
where $\gamma_0$ is the spectral index at the lowest energies, $\gamma_{i-1}$ 
and $\gamma_{i}$ are spectral indices below and above break rigidity 
${\mathcal R_{{\rm br},i}}$, and $s$ describes the smoothness of the break 
which was fixed to be $s=2$ throughout this work. Depending on the assumptions 
and purposes of different models, different numbers of breaks will be assumed. 
Specifically, $n=2$ will be assumed in general, except that the high-energy 
hardening is ascribed to other physical effects ($n=1$ in these cases). 
The spatial distribution of sources of GCRs is parameterized as
\begin{equation}
f(r,z)=\left(\frac{r}{r_\odot}\right)^{\alpha}\exp\left[-\frac
{\beta(r-r_\odot)}{r_\odot}\right]\,\exp\left(-\frac{|z|}{z_s}\right),
\end{equation}
where $r_\odot=8.5$ kpc is the distance from the solar system to the 
Galactic center, $z_s=0.2$ kpc is the scale width of the vertical extension
of sources, $\alpha=1.25$, and $\beta=3.56$ \cite{2011ApJ...729..106T}. 
Unless explicitly stated, we will use the
GALPROP\footnote{https://galprop.stanford.edu} code (version 
56\footnote{A newer version 57 was recently released 
\cite{2022ApJS..262...30P}.}
to calculate the propagation of GCRs \cite{1998ApJ...509..212S}.

To compare with the low-energy measurements in the solar system, we use 
the force-field approximation to account for the solar modulation of GCRs 
\cite{1968ApJ...154.1011G}. More sophisticated models of heliospheric 
propagation exist, e.g., HELMOD \cite{2022AdSpR..70.2636B}, but using 
them is beyond the scope of this paper.

\section{Interpretations of spectral breaks of B/C and B/O}

\subsection{Nested leaky box model}

The leaky-box model, which was popular for the most part of the 20th 
century, is a simplified model with uniform distribution of gas, sources, 
and cosmic rays where the cosmic ray transport in the whole Galaxy is 
described with a single parameter, the escape time 
$\tau_{\rm esc}({\mathcal R})$. Neglecting other processes such as 
the convection, re-acceleration, and fragmentation, the solution of the
propagation equation is as simple as $\psi({\mathcal R})=q({\mathcal R})
\tau_{\rm esc}({\mathcal R})$. An extension of the leaky box model to take into
account the residence and secondary production in dense regions surrounding 
the sources, known as the nested leaky box (NLB) model (denoted as model A), 
was proposed to explain more complicated observational properties of GCRs 
\cite{1973ICRC....1..500C,2010PhRvD..82b3009C,2016ApJ...827..119C}.
In the NLB model, GCRs were accelerated to a power-law spectrum
${\mathcal R}^{-\gamma}$, which diffuse in an energy-dependent way in the 
immediate vicinity of the sources (so-called cocoons), and then enter the 
Galaxy and finally leak to the extragalactic space in an energy-independent 
way. The escape time is assumed to be \cite{2016ApJ...827..119C}
\begin{equation}
\left\{
\begin{array}{llll}
\tau_{\rm esc}^{\rm c}&=&\tau_1 {\mathcal R}^{-\zeta \ln {\mathcal R}}, & {\rm for\ cocoons},\\
\tau_{\rm esc}^{\rm g}&=&\tau_2\equiv{\rm const}, & {\rm for\ Galaxy}.
\end{array}
\right.
\end{equation}

For primary GCRs, the propagated spectrum in cocoons is 
$\psi_{\rm pri}^{\rm c}({\mathcal R})=q({\mathcal R})\tau_{\rm esc}^{\rm c}
\propto {\mathcal R}^{-\gamma-\zeta \ln {\mathcal R}}$. The propagated 
spectrum in the Galaxy is $\psi_{\rm pri}^{\rm g}({\mathcal R})=
[\psi_{\rm pri}^{\rm c}/\tau_{\rm esc}^{\rm c}]\tau_{\rm esc}^{\rm g}
=q({\mathcal R})\tau_{\rm esc}^{\rm g}\propto {\mathcal R}^{-\gamma}$, 
whose spectral shape is the same as the source spectrum.
For secondary particles, there are two components. The one in cocoons
has a spectrum $\psi_{\rm sec}^{\rm c}=\psi_{\rm pri}^{\rm c}\cdot
n_{\rm c}\sigma v\cdot \tau_{\rm esc}^{\rm c}$. This component then
injects into the Galaxy and experiences a further leakage, resulting in
a final spectrum $\psi_{\rm sec}^{\rm c,g}=[\psi_{\rm sec}^{\rm c}/
\tau_{\rm esc}^{\rm c}]\tau_{\rm esc}^{\rm g}$. The other component
is directly produced by GCRs in the Galaxy, whose spectrum is
$\psi_{\rm sec}^{\rm g,g}=\psi_{\rm pri}^{\rm g}\cdot
n_{\rm g}\sigma v\cdot \tau_{\rm esc}^{\rm g}$. The total secondary
spectrum is thus $\psi_{\rm sec}^{\rm g}=\psi_{\rm sec}^{\rm c,g}+
\psi_{\rm sec}^{\rm g,g}=q({\mathcal R})\tau_{\rm esc}^{\rm g}\sigma v
(n_{\rm c}\tau_{\rm esc}^{\rm c}+n_{\rm g}\tau_{\rm esc}^{\rm g})$.
In the above formulae, $n_{\rm c}$ and $n_{\rm g}$ are the gas densities
in cocoons and the Galaxy, $\sigma$ is the production cross section,
and $v$ is the velocity of the GCR particle.

\begin{figure*}[!htb]
\includegraphics[width=\textwidth]{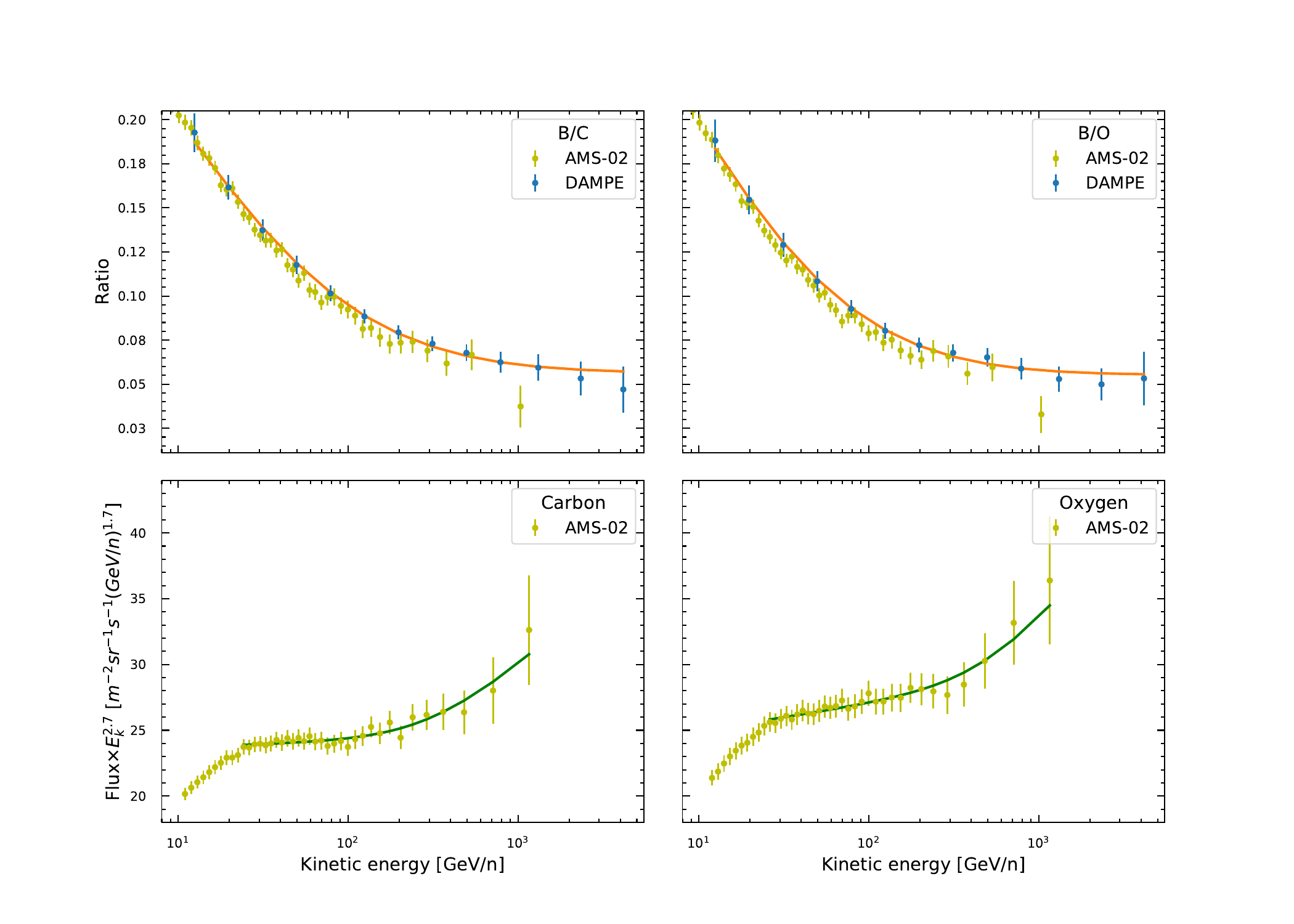}
\caption{B/C and B/O ratios (top panels), and C, O fluxes (bottom panels)
for the NLB model. 
\label{fig:BCO_NLB}}
\end{figure*}

Here we set $n_{\rm c}=1$ H cm$^{-3}$, $n_{\rm g}=0.1$ H cm$^{-3}$,
and derive the other parameters through fitting to the data.
Since some complicated physical effects at low energies (e.g., the 
ionization and Coulomb energy losses) are not included in the NLB model,
we focus on the data-model comparison above a few tens of GeV/n. 
Several experiments found that the spectra of carbon and oxygen nuclei 
are not single power-law, but experience hardening features around a few 
hundred GV \cite{2009BRASP..73..564P,2010ApJ...714L..89A,2017PhRvL.119y1101A,
2020PhRvL.125y1102A}. We therefore assume that the source spectrum is a 
smoothly broken power-law form of rigidity with $n=1$ in Eq.~(\ref{eq:inj}).

Fig.~\ref{fig:BCO_NLB} shows the best-fit B/C, B/O, and C, O fluxes in the NLB 
model, compared with the AMS-02 \cite{2017PhRvL.119y1101A,2018PhRvL.120b1101A} 
and DAMPE \cite{2022DAMPE-BC} data, where the statistical and systematic 
errors of the measurements are added in quadrature.
The fitting parameters are $\gamma_0=2.69\pm0.01$, ${\mathcal R}_{{\rm br},1}
=(533 \pm 292)$ GV, $\gamma_1=2.54\pm0.07$ for C, and $\gamma_0=2.67\pm0.01$,
${\mathcal R}_{{\rm br},1}=(864 \pm 637)$ GV, $\gamma_1=2.51\pm0.11$ for O.
For the secondary-to-primary ratios, we have $\tau_{\rm 1}\sigma=(9.42\pm0.43)
\times10^{-12}$ cm$^{2}$~s, $\tau_{\rm 2}\sigma=(1.88\pm0.06)\times10^{-11}$ 
cm$^{2}$~s, $\zeta=-0.07$ for the B/C ratio, and 
$\tau_{\rm 1}\sigma=(1.04\pm0.06)\times10^{-11}$ cm$^{2}$~s, 
$\tau_{\rm 2}\sigma=(1.84\pm0.06)\times10^{-11}$ cm$^{2}$~s, $\zeta=-0.08$ 
for the B/O ratio. There are many channels to produce boron from fragmentations
of carbon and oxygen \cite{2013ICRC...33..803M,2018PhRvC..98c4611G}.
As an order of magnitude estimate, we take the total fragmentation cross 
section of carbon, $\sim250$ mb, as a reference and obtain $\tau_1\sim1.3$ 
Myr and $\tau_2\sim2.3$ Myr.

In the NLB model, the high-energy behaviors of B/C and B/O asymptotically
approach constants due to the energy-independent leakage in the Milky Way.
This energy-independent leakage predicts a constant dipole 
anisotropy\footnote{The density gradient is ignored in the leaky box model.
Via an analogy with the diffusion model with diffusion coefficient being
scaled to the escape time, the anisotropy in this model was estimated.} 
of GCRs above $\sim$TeV \cite{2010PhRvD..82b3009C}, which is at odds with 
observations. The NLB model is over-simplified, neglecting many important 
processes of propagation of GCRs, and fails to reproduce data in a wide 
energy range. However, the idea that GCRs may propagate differently in 
different regions is important and will be extended to a spatially-dependent 
propagation model or a scenario with confinements and interactions surrounding 
the acceleration sources detailed below.

\subsection{Re-acceleration during propagation}

\begin{figure*}[!htb]
\includegraphics[width=\textwidth]{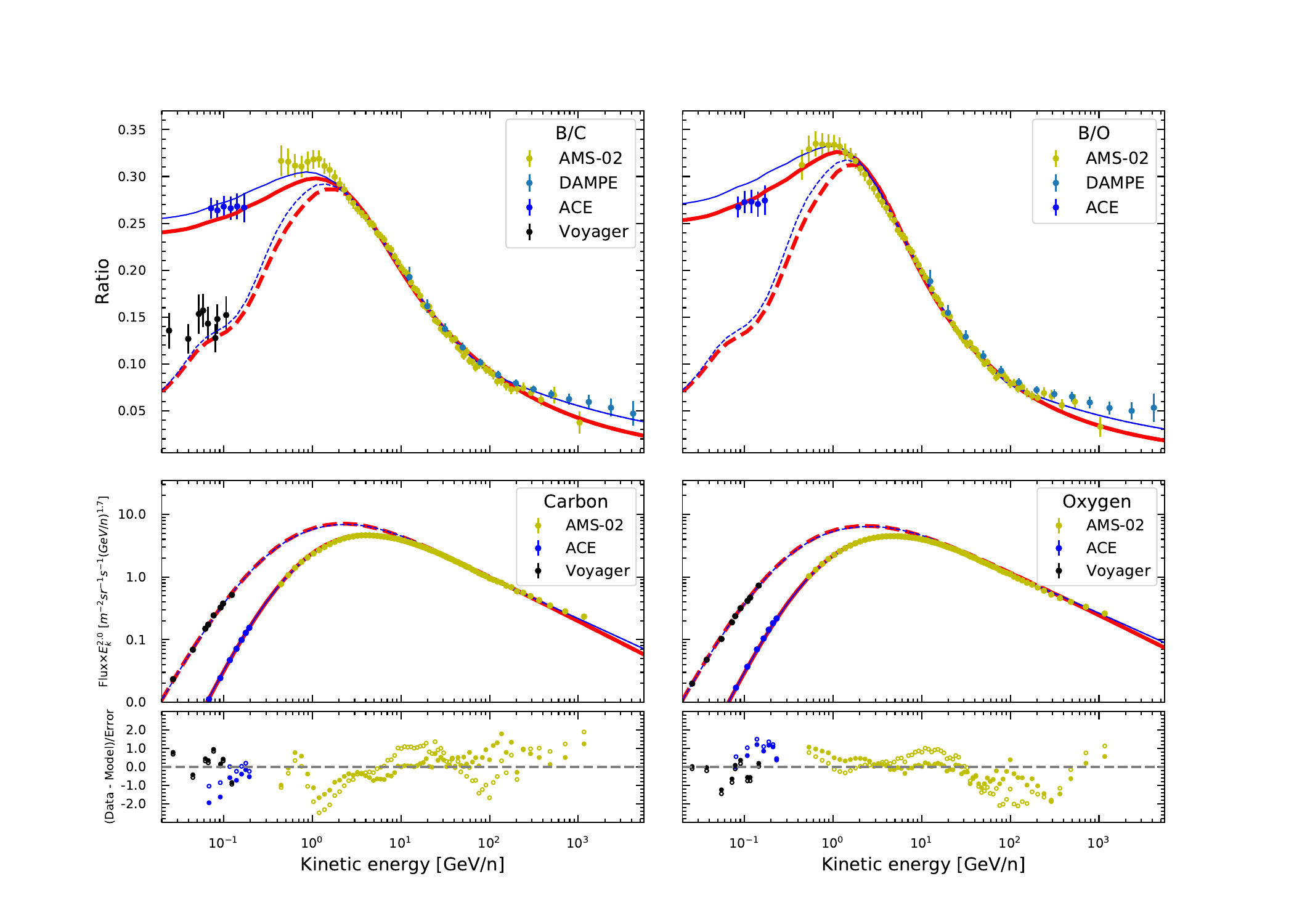}
\caption{B/C and B/O ratios (top panels), and C, O fluxes (bottom panels) 
for the re-acceleration model. Red thick lines show the results for the 
diffusion coefficient of Eq.~(\ref{Dxx}), and blue thin lines show the 
results for the diffusion coefficient with an additional high-energy break. 
Dashed lines are the spectra in the local ISM, and solid lines are modulated 
spectra near the Earth. Sub-panels of bottom ones show the residuals of the 
model fittings to the C and O spectra. The open symbols are for model B and 
filled symbols are for model B$'$.
\label{fig:BCO_Reacc}}
\end{figure*}

GCR particles may get re-accelerated via interactions with randomly 
moving interstellar MHD waves during their propagation process 
\cite{1994ApJ...431..705S}. This stochastic acceleration process 
is usually described by a diffusion in momentum space, with diffusion 
coefficient $D_{pp}$. The re-acceleration results in bump-like spectral 
features of low-energy (less than tens of GeV/n) GCRs, and was shown 
can better explain the peaks of secondary-to-primary ratios 
\cite{2002ApJ...565..280M,2017PhRvD..95h3007Y,2019SCPMA..6249511Y}.
The softer spectra of secondary nuclei experience larger effect of 
re-acceleration, leading to a decrease in the secondary-to-primary ratio 
at low energies.
Fitting to the new measurements of the Li, Be, B, C, and O fluxes by
AMS-02 \cite{2017PhRvL.119y1101A,2018PhRvL.120b1101A} indicates that
the re-acceleration can indeed reproduce well the reported more 
significant hardenings of the secondary family than the primary family
\cite{2020JCAP...11..027Y}. We re-visit the question whether the 
re-acceleration can explain the even stronger hardenings of the B/C 
and B/O ratios measured by DAMPE.

\begin{table}[!htb]
\caption {Data used in the fitting.}
\begin{tabular}{cccc}
\hline \hline
 & Experiment & Time & Ref.\\
\hline
B/C & Voyager & 2012/12-2015/06 & \cite{2016ApJ...831...18C} \\
    & ACE     & 2011/05-2016/05 & \cite{2019SCPMA..6249511Y} \\
    & AMS-02  & 2011/05-2016/05 & \cite{2018PhRvL.120b1101A} \\
    & DAMPE   & 2016/01-2021/12 & \cite{2022DAMPE-BC}\\
\hline
B/O & ACE     & 2011/05-2016/05 & \\
    & AMS-02  & 2011/05-2016/05 & \cite{2018PhRvL.120b1101A} \\
    & DAMPE   & 2016/01-2021/12 & \cite{2022DAMPE-BC} \\
\hline
C \& O & Voyager  & 2012/12-2015/06 & \cite{2016ApJ...831...18C} \\
    & ACE     & 2011/05-2016/05 & \cite{2019SCPMA..6249511Y} \\
    & AMS-02  & 2011/05-2016/05 & \cite{2017PhRvL.119y1101A} \\
\hline
$^{10}$Be/$^9$Be & Voyager & 1977/01-1998/12 & \cite{1999ICRC....3...41L} \\
    & ACE     & 1997/08-1999/04 & \cite{2001ApJ...563..768Y} \\
    & IMP     & 1974/01-1980/05 & \cite{1988SSRv...46..205S} \\
    & Ulysses & 1990/10-1997/12 & \cite{1998ApJ...501L..59C} \\
    & ISOMAX  & 1998/08-1998/08 & \cite{2004ApJ...611..892H} \\
\hline
\hline
\end{tabular}
\label{table:data}
\end{table}

The fitting procedure is similar with Ref.~\cite{2020JCAP...11..027Y}.
We include the Voyager measurements outside the solar system
\cite{2016ApJ...831...18C}, the 5-year AMS-02 secondary-to-primary data 
and 5-year carbon and oxygen data \cite{2017PhRvL.119y1101A,2018PhRvL.120b1101A},
the ACE-CRIS\footnote{http://www.srl.caltech.edu/ACE/ASC/level2/lvl2DATA\_CRIS.html} 
measurements with the same time period of AMS-02, and the DAMPE data.
To reduce the degeneracy between the diffusion coefficient and the halo height, 
the data of $^{10}$Be/$^9$Be from several experiments are also included
\cite{1988SSRv...46..205S,1998ApJ...501L..59C,1999ICRC....3...41L,
2001ApJ...563..768Y,2004ApJ...611..892H}. The data used in the fitting are
summarized in Table \ref{table:data}.

We employ the Markov Chain Monte Carlo (MCMC) method to do the fitting,
using the {\tt emcee} code \cite{2013PASP..125..306F}.
The injection spectrum takes the form of Eq.~(\ref{eq:inj}) with $n=2$.
The best-fit model parameters are given in Table \ref{table:prop} (labelled 
as model B). The red thick lines in Fig.~\ref{fig:BCO_Reacc} show the best-fit
results of the B/C and B/O ratios (top panels), and carbon and oxygen fluxes 
(bottom panels). We can see that the re-acceleration effect can explain partly 
the hardenings of the B/C and B/O ratios, but is not enough to reproduce the 
DAMPE data at high energies.
Therefore, we introduce an additional break for the diffusion coefficient,
i.e., the rigidity-dependence slope becomes $\delta_h$ for ${\mathcal R}>
{\mathcal R}_h$, and re-do the fitting. As shown by the blue thin lines in
Fig.~\ref{fig:BCO_Reacc}, this model (labelled as B$'$) matches both the
ratios and fluxes much better. The best-fit model parameters are also
given in Table \ref{table:prop}. We note that $\gamma_1$ and $\gamma_2$ in 
this model are very close to each other, which means that the hardenings of 
both the primary nuclei and secondary-to-primary ratios are due to the break 
of the diffusion coefficient.

\begin{table}[!htb]
\begin{center}
\caption{Parameters of the models discussed in Sec. III.}
\begin{tabular}{cccccccc}
\hline\hline
   Model & B & B$'$ & D & E & F & G \\ \hline
   $D_0$ ($10^{28}$~cm$^2$~s$^{-1}$) & 6.02 & 3.32  & 8.14  &...& 7.30 & 6.94 \\
   $\delta$                          & 0.40 & 0.46  & 0.60  &...& 0.47 & 0.43 \\
   $z_h$ (kpc)                       & 5.77 & 3.61  & 8.48  &...& 6.23 & 6.29 \\
   $\delta_h$                        & ...  & 0.25  & ...   &...& ... & ... \\
   ${\mathcal R}_{h}$ (GV)           & ...  & 212.5 & ...   &...& ... & ... \\
   $v_A$ (km s$^{-1}$)               & 30.0 & 22.4  & 10.3  &...& 35.4 & 32.5 \\
   $\eta$                            &$-0.10$ &$-0.61$ & $-0.42$&...&$-0.27$&$-0.33$\\
   $\xi$                             & ...  & ...   & 0.1  &...& ... & ... \\
   $\xi_{\delta}$                    & ...  & ...   & 0.02  &...& ... & ... \\
   $h$ (kpc)                         & ...  & ...   & 0.41  &...& ... & ... \\
   $\tau$ (Myr)                      & ...  & ...   & ...   &...& 0.46 & 0.24 \\
   ${\mathcal R}_{\rm ac}$ (GV)      & ...  & ...   & ...   &...& ...  & $4.0\times10^3$ \\
\hline
   $\gamma_0$                        & 0.19 & 0.41  & 0.13  &0.85& 0.46 & 0.45 \\
   $\gamma_1$                        & 2.36 & 2.35  & 2.38  &2.37& 2.31 & 2.34 \\
   $\gamma_2$                        & 2.34 & 2.42  & ...   &...& 2.15 & 2.18 \\
   ${\mathcal R}_{\rm br,1}$ (GV)    & 0.93 & 1.02  & 1.03  &1.75& 1.05 & 1.05 \\
   ${\mathcal R}_{\rm br,2}$ (GV)    & 243.2& 142.3 & ...   &...& 467.0 & 467.0\\
\hline
   $\phi$ (GV)                       & 0.71 & 0.69  & 0.71  &0.68& 0.71 & 0.69 \\
\hline \hline
\end{tabular}
\label{table:prop}
\end{center}
\end{table}

\subsection{Re-acceleration by a nearby source}

Malkov and Moskalenko proposed recently that the re-acceleration of GCRs
by a close star, such as Epsilon Eridani, can well explain the observed 
bump structures of GCRs \cite{2021ApJ...911..151M,2022ApJ...933...78M}. 
This scenario predicts hardenings of both primary and secondary nuclei 
which were re-accelerated simultaneously. Different from the stachastic 
acceleration in the ISM described in the previous sub-section (where the 
impacts are mainly at low energies, e.g., for rigidity below tens 
of GV), this model mainly affects the GCR spectra above TV rigidities 
since low-energy re-accelerated particles are convected with the ISM and 
do not reach the solar system. It thus gives bump-like features of the 
GCR spectra around 10 TV, explaining both the hardenings and consequent 
softenings of the spectra.

The re-accelerated spectrum of a power-law background spectrum can be
described as \cite{2022ApJ...933...78M}
\begin{equation}
f({\mathcal R})=q({\mathcal R})\left[1+\frac{\gamma+2}{\lambda-\gamma}\exp
\left(-\sqrt{\frac{{\mathcal R}_0}{{\mathcal R}}}-
\sqrt{\frac{{\mathcal R}}{{\mathcal R}_L}}\right)\right],
\end{equation}
where $q({\mathcal R})=q_0 {\mathcal R}^{-\gamma}$ is the background
GCR spectrum, ${\mathcal R}_0$ is a characteristic rigidity of the 
re-accelerated bump depending on the shock properties, ${\mathcal R}_L$
is the rigidity cutoff associated with the lateral losses,
$\lambda=(r+2)/(r-1)$ with $r$ being the shock compression ratio.
For the ``realistic'' model of Ref.~\cite{2022ApJ...933...78M}, 
the fitting to the proton spectrum gives ${\mathcal R}_0=5.9$ TV,
${\mathcal R}_L=224$ TV, $q=4.2$. The same parameters can be applied
to other nuclei such as helium, boron, and carbon, and good consistencies
with the data were shown \cite{2022ApJ...933...78M}.

\begin{figure*}[!htb]
\includegraphics[width=\textwidth]{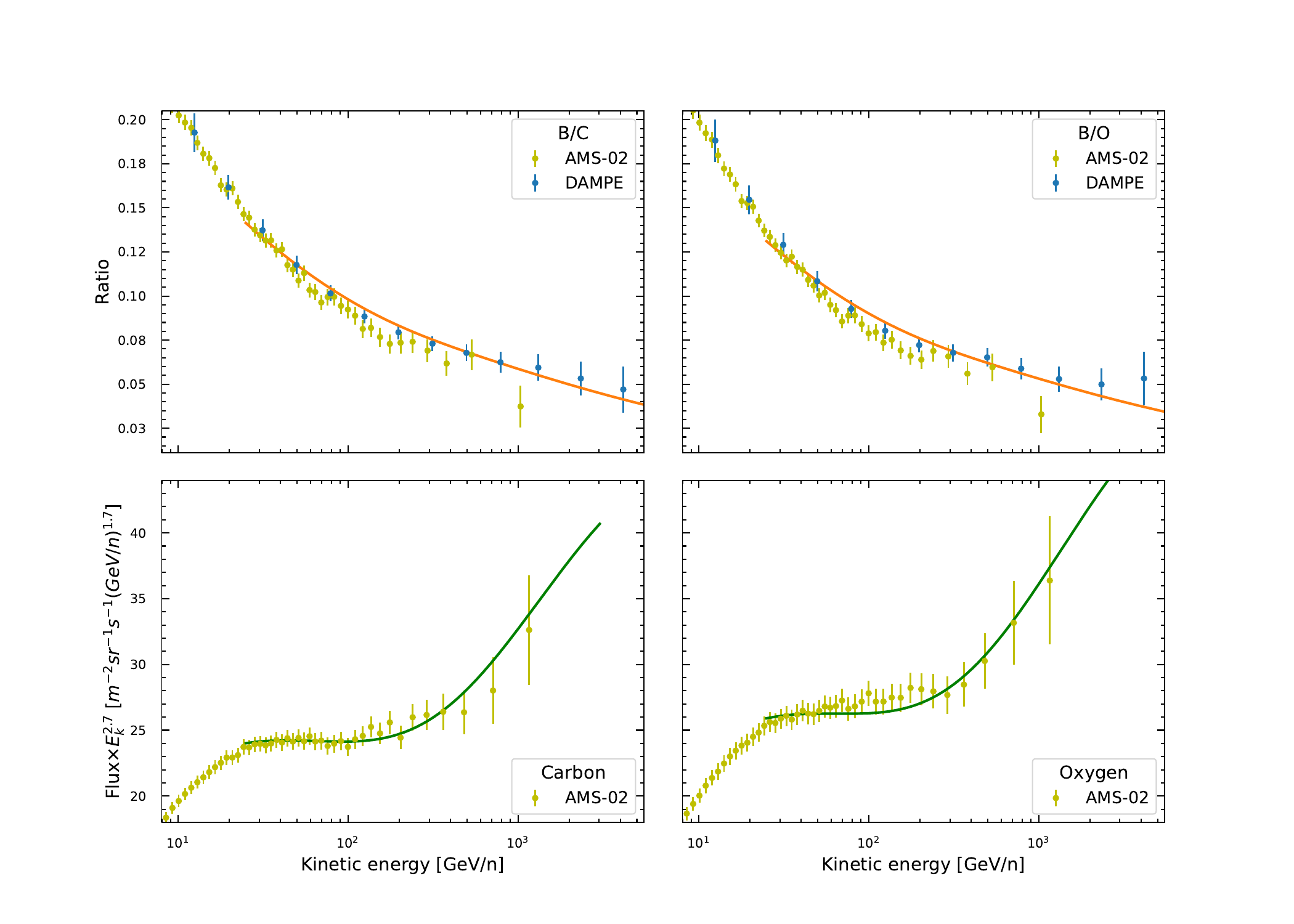}
\caption{B/C and B/O ratios (top panels), and C, O fluxes (bottom panels)
for the model with re-acceleration by a nearby source.
\label{fig:BCO_RM}} 
\end{figure*} 

Fig.~\ref{fig:BCO_RM} shows the comparison between the model predictions
and the new measurements of B/C, B/O, and C, O fluxes of this scenario
(model C), adopting the same shock parameters as in 
Ref.~\cite{2022ApJ...933...78M}. The spectral parameters of background 
GCRs are slightly adjusted, i.e., $\gamma=2.76$ for carbon, $2.75$ for 
oxygen, and $3.04$ for boron. While improvement of the data-model match 
can be expected given a re-fitting of all the data, we find that this 
simple model reproduces the current measurements reasonably well.

\subsection{Spatially-dependent propagation}

In the conventional propagation model, the diffusion coefficient is
assumed to be homogeneous throughout the Milky Way. While this model
can explain most of the GCR spectra and all-sky diffuse $\gamma$-rays
\cite{2007ARNPS..57..285S,2012ApJ...750....3A}, 
recent observations indicate that the GCR
propagation is likely spatially-dependent. Very high energy extended
$\gamma$-ray halos around a few pulsars observed by HAWC and LHAASO
suggest that particles propagate very slowly in the ISM surrounding 
pulsars \cite{2017Sci...358..911A,2021PhRvL.126x1103A}. Together with the 
diffusion coefficient inferred from the secondary-to-primary ratios from 
GCR direct measurements, the propagation of GCRs could be inhomogeneous 
--- slow in the Galactic disk (or the vicinities of sources) and fast 
in the halo \cite{2017PhRvD..96j3013H,2018ApJ...863...30F}.
The spatially-dependent propagation model was also employed to explain
the hundreds of GV hardenings of GCR spectra and the high energy excess
of the Fermi-LAT diffuse $\gamma$-rays \cite{2012ApJ...752L..13T,
2018PhRvD..97f3008G,2022FrPhy..1744501Q}. The Bayesian analysis to derive the 
propagation parameters within the spatially-dependent propagation framework 
has also been carried out \cite{2016PhRvD..94l3007F,2021PhRvD.104l3001Z}. 

We parameterize the spatial diffusion coefficient $D_{xx}$ as
\begin{equation}
D_{xx}({\mathcal R},z)=aD_0\beta^{\eta}\left(\frac{{\mathcal R}}
{{\mathcal R}_0}\right)^{b\delta},
\end{equation}
with $a=\xi+(1-\xi)[1-\exp(-z^2/2h^2)]$, $b=\xi_{\delta}+(1-\xi_{\delta})
[1-\exp(-z^2/2h^2)]$, where $h$ is the characteristic thickness of the
slow-diffusion disk region, $\xi$ and $\xi_{\delta}$ are suppression
factors of the diffusion coefficient and its rigidity-dependent slope
in the disk region compared with those in the fast-diffusion halo region.
The reference rigidity ${\mathcal R}_0$ is fixed to be 4 GV.
Note here we ignore the possible $R$-dependence of the diffusion
coefficient \cite{2016ApJ...819...54G}. For $z\gg h$ (fast-diffusion halo),
we have $a=b=1$, and for $z\to 0$ (slow-diffusion disk), we have $a\to \xi$,
$b\to\xi_{\delta}$.

\begin{figure*}[!htb]
\includegraphics[width=\textwidth]{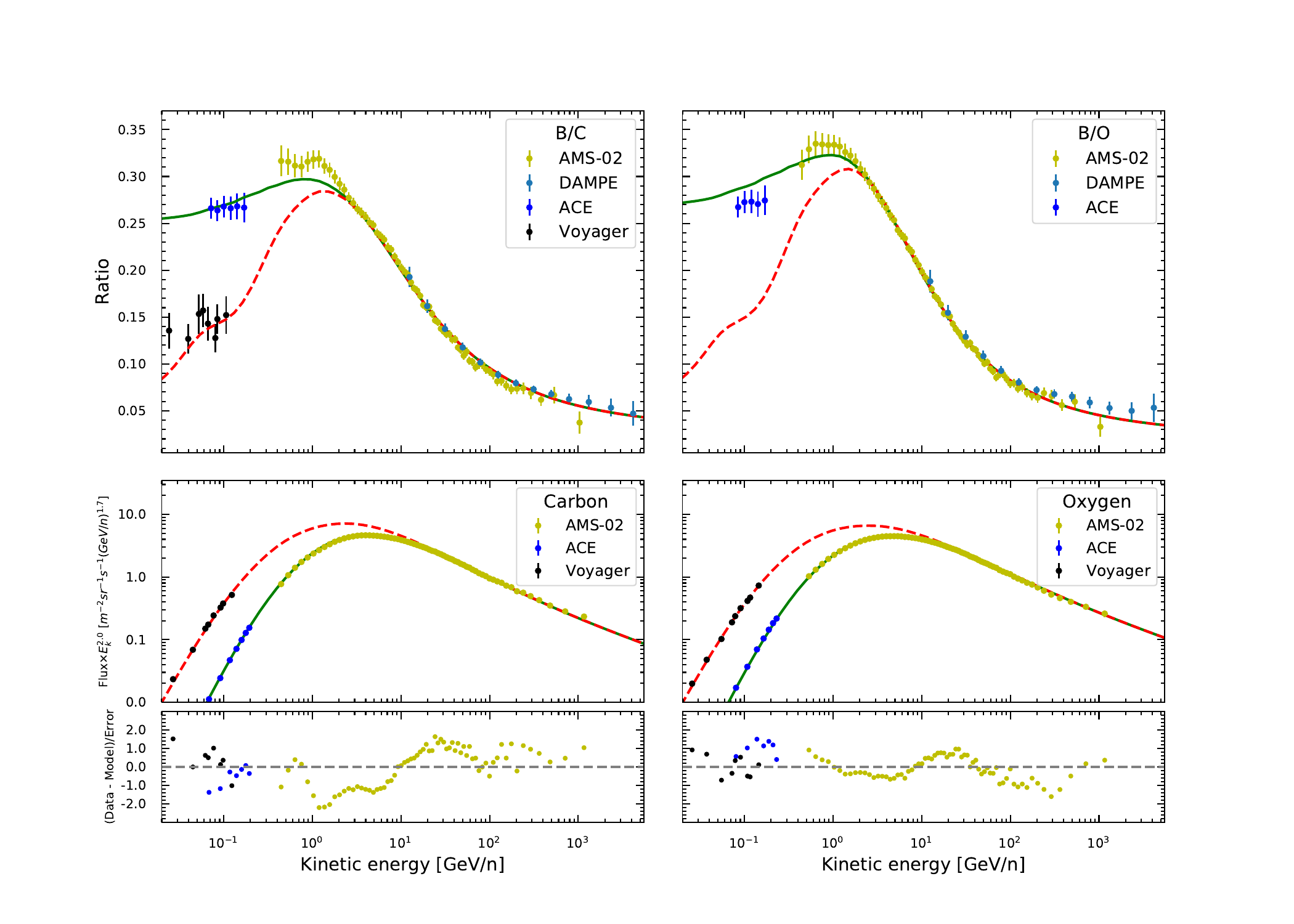}
\caption{B/C and B/O ratios (top panels), and C, O fluxes (bottom panels)
for the spatially-dependent propagation model. Dashed lines are the spectra 
in the local ISM, and solid lines are modulated spectra near the Earth.
Sub-panels of bottom ones show the residuals of the model fittings 
to the C and O spectra.
\label{fig:BCO_SDP}}
\end{figure*}

We fit the model parameters using the same data sets and fitting procedure
as in Sec. III B. In this model the hardenings at hundreds of GV are primarily
ascribed to the spatially-dependent propagation effect. Hence for the injection
spectrum we assume $n=1$, where the break occurs around GV to account for the
low energy data. The results are shown in Fig.~\ref{fig:BCO_SDP} and
Table \ref{table:prop}. It can be seen that the model reproduces the data 
very well. We note that the diffusion coefficient in the disk is smaller by
about three orders of magnitude than that in the halo, for a rigidity of 
$\sim100$ TV, which is consistent with the results inferred from pulsar
halo observations \cite{2017Sci...358..911A,2021PhRvL.126x1103A}. 
The parameter $\xi_{\delta}=0.03$ is also close to the result ($\sim 0$) 
obtained in Ref.~\cite{2021PhRvD.104l3001Z}. This nearly rigidity-independent
diffusion coefficient in the disk may be tested by future observations of
the energy-dependent morphologies of pulsar halos.

\subsection{Self-generated turbulence model}

The GCR flows can induce MHD waves of the background plasma through 
the streaming instability \cite{1971ApJ...170..265S}, leading to         
self-confinement of GCRs around such waves. This nonlinear effect 
makes GCR propagation couples with the interstellar waves, resulting
in changes of the momentum dependence and spatial dependence of the
diffusion coefficient, which was employed to explain the hardening
feature \cite{2012PhRvL.109f1101B} and radial distribution 
\cite{2016MNRAS.462L..88R} of GCRs. In this scenario, the break of GCR
spectra around ${\mathcal R}\sim 10$ GV is due to the transition of GCR
propagation from advection to the regime dominated by diffusion in
self-generated turbulence, and the hardening around ${\mathcal R}\sim200$ 
GV is due to a further transition of the diffusion from self-generated
turbulence to externally generated turbulence \cite{2012PhRvL.109f1101B}.

\begin{figure*}[!htb]
\includegraphics[width=\textwidth]{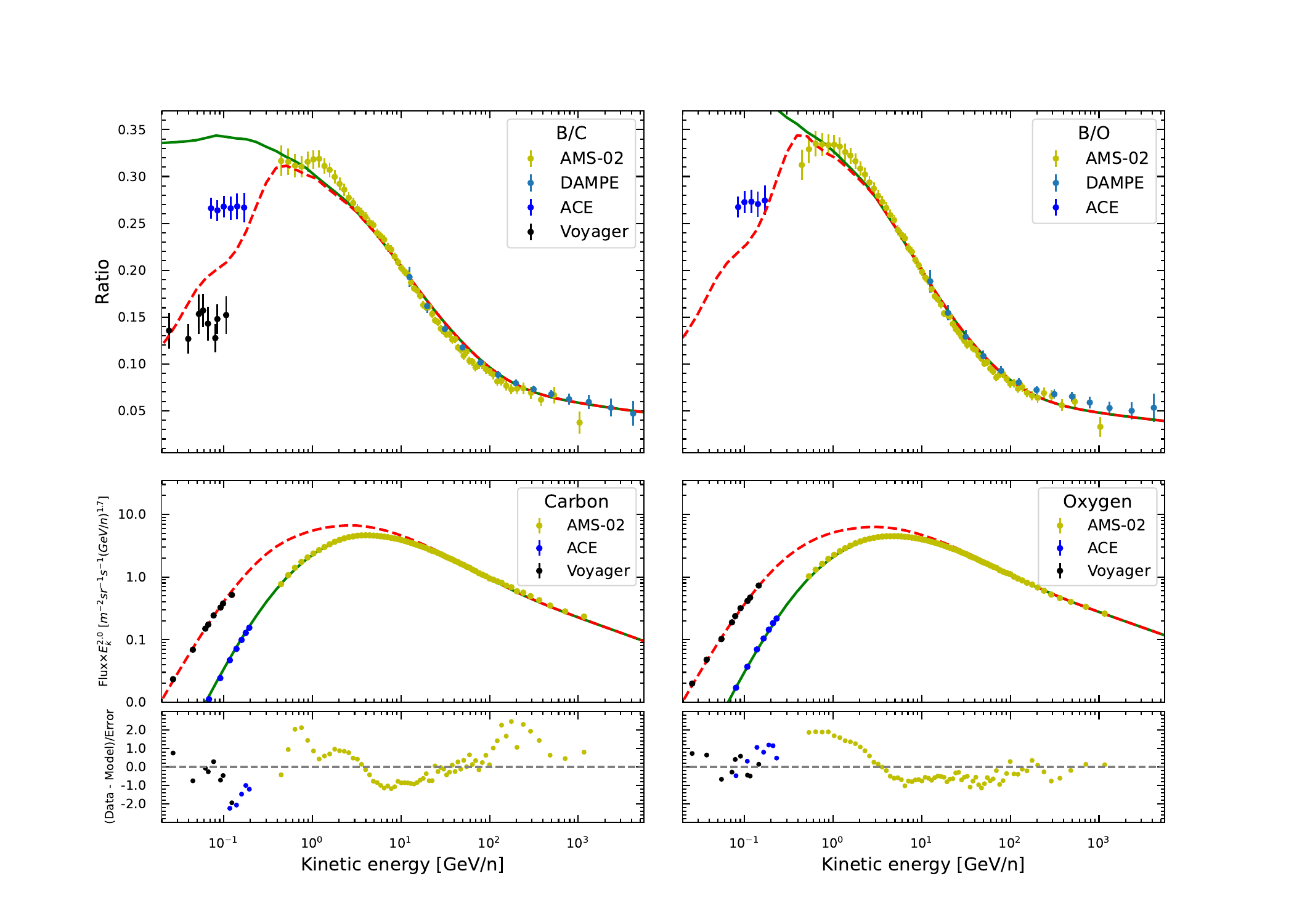}
\caption{B/C and B/O ratios (top panels), and C, O fluxes (bottom panels)
for the Self-generated turbulence model. Dashed lines are the spectra 
in the local ISM, and solid lines are modulated spectra near the Earth.
Sub-panels of bottom ones show the residuals of the model fittings 
to the C and O spectra.
\label{fig:BCO_Selfgen}}
\end{figure*}

A self-consistent treatment of this problem needs to solve the GCR 
transportation and the MHD wave evolution simultaneously. For the purpose 
of illustration, we simply adopt the momentum-dependence of the diffusion 
coefficient given in Ref.~\cite{2012PhRvL.109f1101B} and neglects its 
spatial variation. At low energies, the slope of the diffusion coefficient 
was found to be $\sim{\mathcal R}^{0.7}$, which aymptotically changes to
$\sim{\mathcal R}^{1/3}$ at high energies. To better fit the data, we 
adjust the diffusion coefficient of Ref.~\cite{2012PhRvL.109f1101B} 
through multiplying a factor $a({\mathcal R}/{\rm GV})^b$. The injection 
spectrum is again Eq.~(\ref{eq:inj}), with $n=1$. We find that $a=1.72$ 
and $b=-0.23$ can fit the data above a few GV relatively well, as shown 
in Fig.~\ref{fig:BCO_Selfgen}. The low-energy parts of the B/C and B/O
ratios are over-predicted by this model. A modification of the
velocity-dependence of the diffusion coefficient as in Eq.~(\ref{Dxx})
may be helpful in improving the fitting.

\subsection{Secondary production at sources}

In the standard model, secondary particles are produced by inelastic
interactions of primary GCRs with the ISM during the propagation process
in the Milky Way \cite{2007ARNPS..57..285S}. It is possible that the same
interactions occur in the vicinities of GCR sources, particularly in the
case that there are dense molecular clouds surrounding the sources.
Such interactions were proposed to explain the positron excess
\cite{2009PhRvD..80f3003F,2016PTEP.2016b1E01K,2016PhRvD..94f3006M,
2017PhRvD..96b3006L,2019PhRvD.100f3020Y},
and the ultra-high-energy diffuse $\gamma$-ray emission measured by
Tibet AS$\gamma$ \cite{2021PhRvL.126n1101A,2022PhRvD.105b3002Z}.
To account for the positron excess, this new secondary component is
required to be close to the Earth, and the time-dependent propagation
is employed to suppress low-energy particles and to account for the
measured high-energy excess of positrons. On the other hand, the
general secondary interactions around sources can contribute to the
high-energy positrons, but cannot fit the data nicely, and additional
nearby source(s) (as also required by the primary GCR spectral
features and anisotropies) is assumed \cite{2022PhRvD.105b3002Z}.

Such interactions should imprint on the B/C and B/O ratios
\cite{2022PhRvD.105b3002Z,2023JCAP...02..007Z}. Different from 
Refs.~\cite{2022PhRvD.105b3002Z,2023JCAP...02..007Z}, we investigate the 
effects from secondary interactions in both the ISM and the vicinities of 
the sources, without assuming the nearby source component. The hardenings 
of primary GCR spectra are assumed to be a source injection effect
\cite{2012ApJ...752...68V,2020ApJS..250...27B}. The injection 
spectrum takes the form of Eq.~(\ref{eq:inj}), with $n=2$. 
The secondary source function can be written as
\begin{eqnarray}
q_{{\rm sec},j} & = & \sum_{i} (n_{\rm H} \sigma_{i+{\rm H}\to j} +
n_{\rm He} \sigma_{i+{\rm He}\to j} )v_i \nonumber\\
& \times & \left[\psi_i({\mathcal R}) + q_i({\mathcal R}) \tau\right],
\label{eq:qsec}
\end{eqnarray}
where $i$ represents any species that can fragment into secondary
particle $j$, and $v_i$ is the velocity of the parent particle.
The first term in the bracket represents the equilibrium density
of parent GCRs after the propagation, while the second term is the
contribution from secondary interactions around sources in which $q_i$
is the injection source function, and $\tau$ is the interaction time scale. 

\begin{figure*}[!htb]
\includegraphics[width=\textwidth]{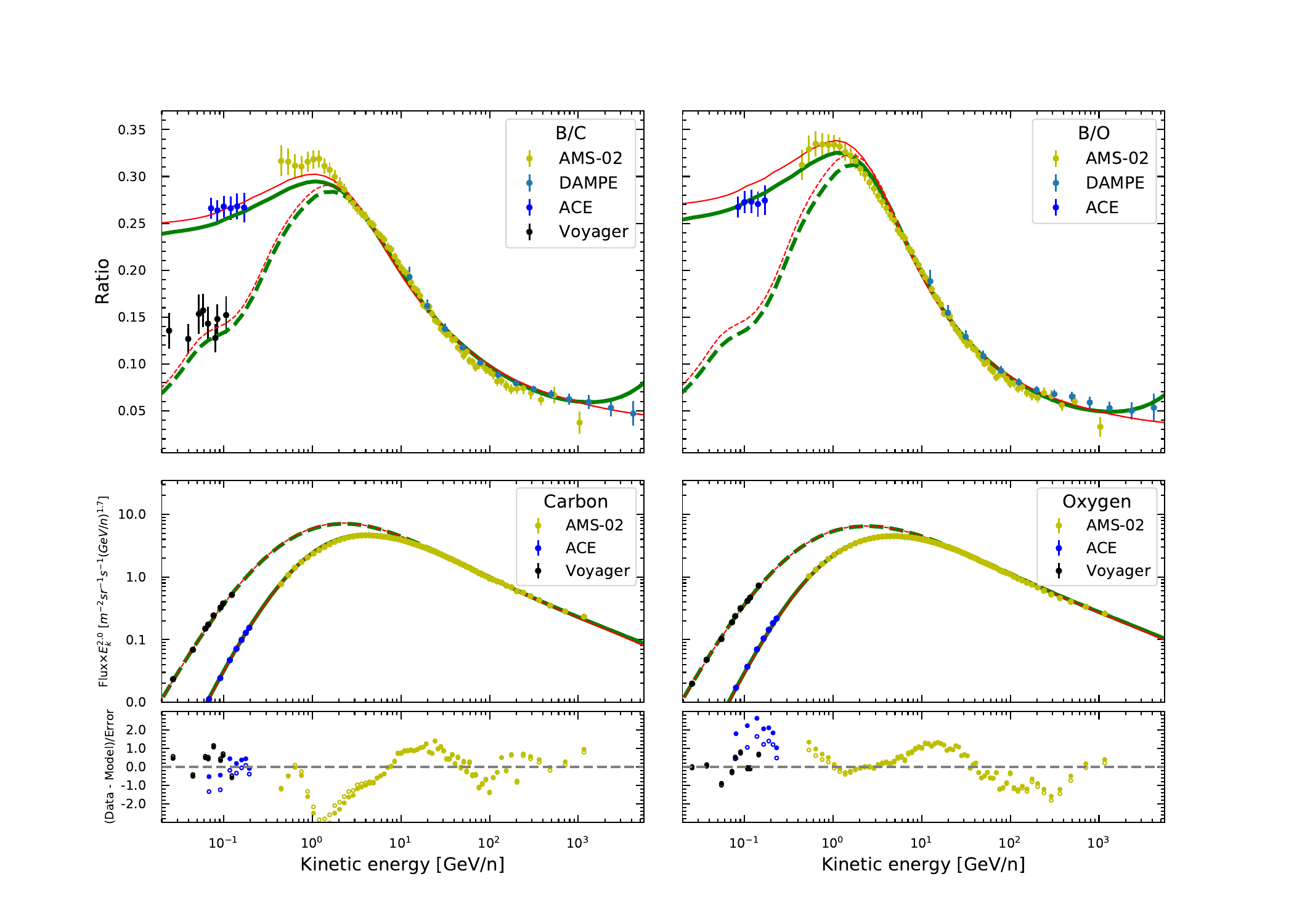}
\caption{B/C and B/O ratios (top panels), and C, O fluxes (bottom panels)
for the models with secondary production (model F; red thin lines) and 
acceleration (model G; green thick lines) at source. Dashed lines are the 
spectra in the local ISM, and solid lines are modulated spectra near the Earth. 
Sub-panels of bottom ones show the residuals of the model fittings 
to the C and O spectra. The filled symbols are for model F and open symbols 
are for model G.
\label{fig:BCO_Secgen}}
\end{figure*}

We insert the $q_i({\mathcal R}) \tau$ term in the routine to calculate 
the secondary source function in GALPROP, and calculate its propagation 
simultaneously with the conventional secondary particles. The propagation 
and source parameters tuned to match the data are given in 
Table~\ref{table:prop}. The red thin lines in Fig.~\ref{fig:BCO_Secgen}
show the results for the model predictions (labelled as model F), compared 
with the data. The $q_i({\mathcal R}) \tau$ term results in a flat 
secondary-to-primary ratio above a few GeV/n, while the $\psi_i$ term gives 
a decreasing ratio. The sum of these two components naturally explains the 
hardenings of the B/C and B/O ratios. Note that the results of this scenario 
are similar with the NLB model, but with different physical meanings. In the 
NLB model, a rigidity-dependent escape is assumed in the vicinities of the
sources, and a constant escape is assumed for the Milky Way propagation.
Here, on the contrary, the Milky Way propagation is rigidity-dependent,
and inside the source regions, confinements of both primary and secondary 
nuclei are assumed. The interaction time in this case is found to be 0.46 Myr. 
This time scale is too long compared with the typical life time of supernova 
remnants ($<10^5$ yr; \cite{2019ApJ...874...50Z}). If the sources are 
associated with molecular clouds in general, the required interaction time 
can be shorter.

\subsection{Secondary production and acceleration at sources}

Secondary particles generated close to the accelerating sources may have
chance to be accelerated by the shocks of the sources, resulting in harder
spectra of secondary particles than the primary particles, which can 
explain the positron excess \cite{2003A&A...410..189B,2009PhRvL.103e1104B,
2009PhRvD..80l3017A}. The secondary-to-primary ratios of nuclei were shown 
to be sensitive probes of this model 
\cite{2009PhRvL.103h1104M,2014PhRvD..89d3013C,2021PhRvD.104j3029M}.

Assuming that the primary particles accelerated by the source have a
power-law spectrum of ${\mathcal R}^{-\gamma}$, the spectrum of secondary
particles can be expressed as the sum of two power-laws: one with
approximate ${\mathcal R}^{-\gamma}$ spectrum describing the component
advected away from the shock, and the other with approximate
${\mathcal R}^{-\gamma+\delta}$ is the component subject to additional
acceleration by the shock \cite{2009PhRvD..80l3017A,2009PhRvL.103h1103B}.
Here $\delta$ is the slope of the rigidity-dependence of the diffusion
coefficient around the shock. For Bohm-like diffusion which corresponds 
to the regime that the mean free path of a particle is of the order of the 
gyroradius and is usually assumed for diffusive shock acceleration with a 
possible fudge factor (e.g., \cite{2009PhRvL.103e1104B,2009PhRvD..80l3017A}), 
$\delta=1$. Therefore, we multiply a factor\footnote{The condition that the 
number of secondary particles being accelerated should not exceed the total 
number of secondaries limits the growth of the accelerated term. This 
affects the energy spectrum of secondary particles at a break rigidity 
above which the spectrum returns to ${\mathcal R}^{-\gamma}$ 
\cite{2009PhRvD..80l3017A}. The break may occur at rigidities too high
to be probed by the current data, and thus we do not consider it here.}
$(1+{\mathcal R}/{\mathcal R}_{\rm ac})$ to the propagated component of 
secondary particles produced at source. Here ${\mathcal R}_{\rm ac}$ is 
the characteristic rigidity that accelerated secondary particles become 
important. The parameter ${\mathcal R}_{\rm ac}$ depends on the diffusion 
coefficient and shock parameters, which can be derived via fitting to the 
B/C and B/O data. 

The results for B/C, B/O, C and O fluxes for $\tau=0.24$ Myr and 
${\mathcal R}_{\rm ac}=4$ TV are shown by green thick lines of 
Fig.~\ref{fig:BCO_Secgen}. Other parameters of this model (model G) are
given in Table \ref{table:prop}. Rising behaviors of the B/C and B/O ratios 
are predicted in this scenario. Since we do not observe such a rising in the 
DAMPE data, ${\mathcal R}_{\rm ac}$ is constrained to be higher than previous 
works \cite{2009PhRvL.103h1104M,2014PhRvD..89d3013C,2021PhRvD.104j3029M}.
The parameter ${\mathcal R}_{\rm ac}$ may be even higher if we require that 
there is no significant rising of the B/C ratio above 2 TeV/n. Then this 
model further degenerates with model F in the previous sub-section.

\section{Implications on other secondary particles}

\begin{figure*}[!htb]
\includegraphics[width=\textwidth]{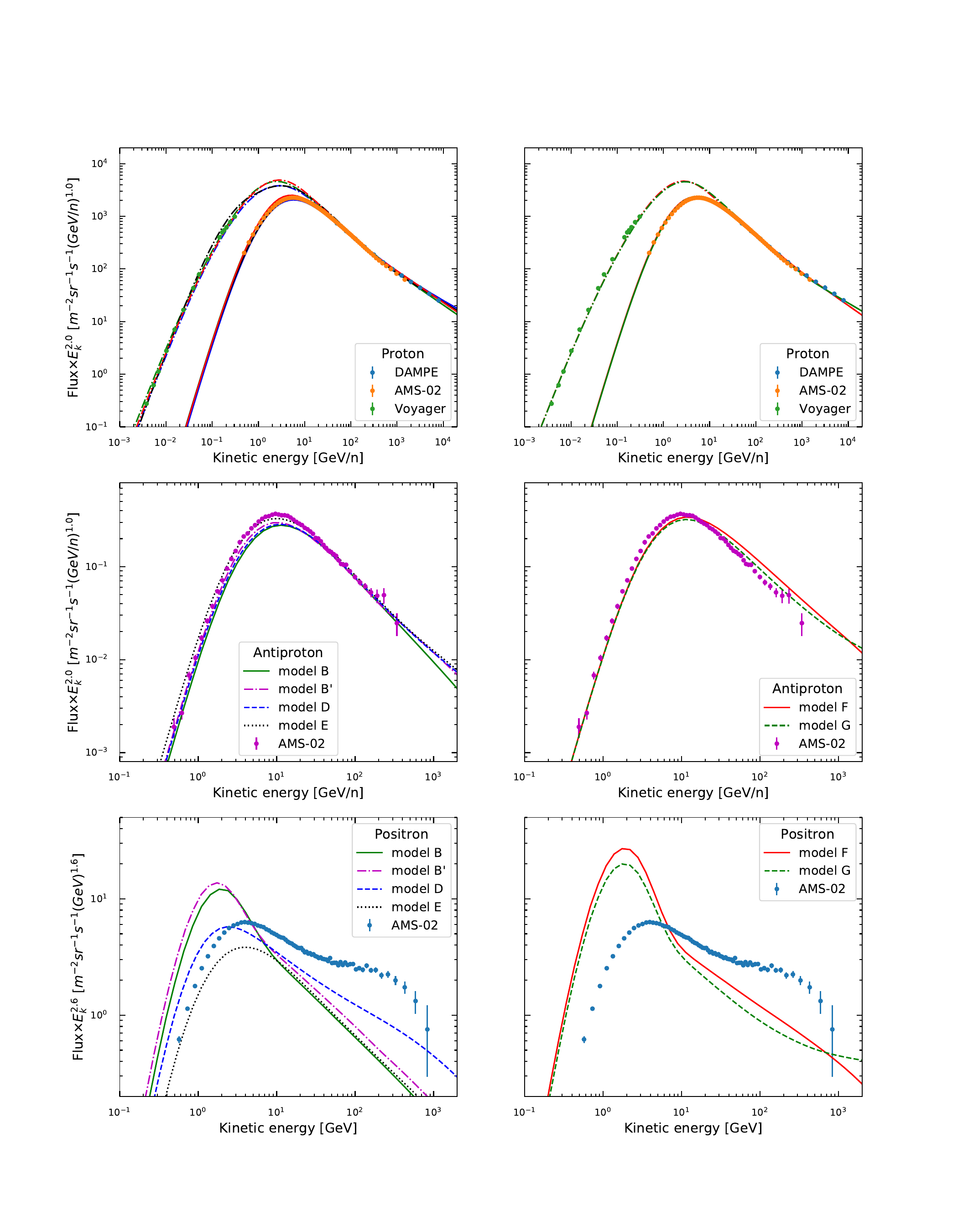}
\caption{Spectra of protons (top panels), antiprotons (middle panels), 
and positrons (bottom panels) for the model predictions, compared with
the data \cite{2016ApJ...831...18C,2015PhRvL.114q1103A,2019SciA....5.3793A,
2016PhRvL.117i1103A,2019PhRvL.122d1102A}. The left panels show the predictions 
of models B, B$'$, D, E, and the right panels show the predictions of models 
F and G. The solar modulation potential is about $0.7$ GV for all species.
\label{fig:pbar_pos}}
\end{figure*}

The effects on the production of secondary boron nuclei are also expected 
to imprint on other secondary particles, such as antiprotons and positrons.
Of particular interests is that antiprotons and positrons are widely employed
to search for dark matter annihilation or decay. It is thus necessary to 
investigate how the change of the boron production or propagation affects 
the predictions of secondary antiprotons and positrons.

Using the propagation parameters given in Table \ref{table:prop}, we adjust
slightly the injection spectral parameters to match the wide-band measurements
of protons and helium nuclei by Voyager, AMS-02, and DAMPE
\cite{2016ApJ...831...18C,2015PhRvL.114q1103A,2019SciA....5.3793A,
2017PhRvL.119y1101A,2021PhRvL.126t1102A}, and then calculate the spectra of 
antiprotons and positrons. The results are shown in Fig.~\ref{fig:pbar_pos}.
The left panels correspond to models B, B$'$, D, and E which represent the
class driven by propagation effects, while the right panels 
correspond to models F and G which represent the class driven by 
source effects.

As can be seen from the plot, the propagation effect will in general result
in a hardening of the antiproton spectrum above a few hundred GV rigidity,
similar with the B/C and B/O ratios. The source effect is somehow different,
due to different kinematics of the production of boron nuclei, antiprotons,
and positrons. The inelasticity parameter, describing the fraction of energy
of the parent particle carried away by the secondary particle, is about 1
for boron nuclei, 0.17 for antiprotons, and 0.05 for positrons. Therefore
we see that the effect of secondary production at sources start to appear
at lower energy for antiprotons than that for boron, which results in 
slight excess of antiprotons above 20 GeV. At low energies, we note 
that most of the model calculations are lower than measured antiproton 
fluxes\footnote{See for example, a dark matter annihilation explanation
of this potential excess \cite{2017PhRvL.118s1101C,2022PhRvL.129i1802F}.}. 
This is consistent with previous studies that the model with significant 
re-acceleration would under-predict low-energy antiprotons
\cite{2003ApJ...586.1050M,2017PhRvD..95h3007Y}, likely due to a smaller 
$\delta$ value in such models. If the re-acceleration is weaker 
(model B$'$) or there is no re-acceleration (model E), the low-energy
antiproton deficit is less significant. We should bear in mind that 
uncertainties of the inelastic hadronic interaction cross section to produce 
antiprotons may need to be considered when quantitative comparisons are 
performed.

As for positrons, all the models predict lower positron fluxes than the
AMS-02 measurements \cite{2019PhRvL.122d1102A} above $\sim10$ GeV, indicating
that additional positron sources are required. For $E\lesssim10$ GeV,
the re-acceleration models typically over-predict positrons, consistent 
with previous studies \cite{2002ApJ...565..280M,2017PhRvD..95h3007Y}.
For models D and E, less prominent bumps are given due to a smaller 
re-acceleration term. We note that the spatially-dependent propagation
model (model D) predicts harder positron spectrum than other models. 
This is mainly because the competition between diffusion and cooling is 
different for the disk and halo regions in this model. In the disk the
diffusion coefficient is very small, and the cooling effect dominates
the propagation, which results in a soft positron spectrum. On the other
hand, the diffusion becomes much more important in the halo, which results
in a harder spectrum. The hard spectrum thus reflects the fast diffusion
in the Galactic halo.

\section{Conclusion}

In this work we study a series of revised models of the standard GCR
propagation in light of new measurements of B/C and B/O ratios by DAMPE
which firmly established hardenings of both ratios around 100 GeV/n
\cite{2022DAMPE-BC}. These models, as detailed in Sec. III, rely on
different physical assumptions and turn out to modify the propagation 
process (models A, B, B$'$, C, D, E) or production of secondary particles 
(models F, G) based on the conventional paradigm. Our conclusion can be 
summarized as follows.

\begin{itemize}

\item About half of the models (B$'$, D, F, and G) can properly reproduce 
the B/C and B/O ratios as well as C, O fluxes in a wide energy range from 
0.01 GeV/n to 5 TeV/n. For the NLB model (A) and re-acceleration by a local
shock model (C) we only compare them with the data above 25 GeV/n, and for 
the self-generated turbulence model (E) we do not tune the results below 
$\sim3$ GeV/n. The re-acceleration model (B) is not enough to give the 
prominent hardenings of the B/C and B/O ratios revealed by DAMPE.
The model of production and acceleration of secondary particles at source 
(G) predicts rising behavior of the ratios at high energies, which is not 
shown by the current data.

\item The models with significant re-acceleration (a large $v_A$; B, B$'$,
F, G) under-predict low energy antiprotons, but over-predict low-energy 
positrons. 

\item The models with secondary production at sources (without and with
acceleration) over-predict high-energy antiprotons. 

\item For all the models discussed in this work, high-energy positron
excess still exists, which requires additional sources of positrons.

\end{itemize}

It is shown that the new measurements of GCR spectra and ratios in recent 
years can indeed provide important constraints on the propagation and
interaction of GCRs in the Milky Way. Additional tests of these models may 
include the anisotropies of GCRs and wide-band diffuse $\gamma$-rays.
In addition, the uncertainties from nuclear and particle physics become
more and more prominent when making precise comparisons between model
predictions and the astroparticle data, particularly for the search for
dark matter.

\acknowledgments
This work is supported by the National Key Research and Development Program of China 
(No. 2021YFA0718404), the National Natural Science Foundation of China (Nos. 12220101003, 
12103094), and the Project for Young Scientists in Basic Research of Chinese Academy of 
Sciences (No. YSBR-061). 
The calculation is partially done on the Cosmology Simulation Database (CSD) of the 
National Basic Science Data Center (NBSDC-DB-10).

\bibliographystyle{apsrev4-1}
\bibliography{refs}

\end{document}